# Reconstructing phase-resolved hysteresis loops from first-order reversal curves

Dustin A. Gilbert,[1,2,*] Peyton D. Murray,[3] Julius De Rojas,[3] Randy K. Dumas,[4] Joseph E. Davies,[5] Kai Liu[3,6]

[1] *Department of Materials Science and Engineering, University of Tennessee, Knoxville, Tennessee 37919*

[2] *Department of Physics, University of Tennessee, Knoxville, Tennessee 37919*

[3] *Physics Department, University of California, Davis, CA 95616*

[4] *Quantum Design, Inc., San Diego, California 92121*

[5] *Advanced Technology Group, NVE Corp., Eden Prairie, MN, 55344*

[6] *Department of Physics, Georgetown University, Washington, D.C. 20057*

* dagilbert@utk.edu

## Abstract

The first order reversal curve (FORC) method is a magnetometry based technique used to capture nanoscale magnetic phase separation and interactions with macroscopic measurements using minor hysteresis loop analysis. This makes the FORC technique a powerful tool in the analysis of complex systems which cannot be effectively probed using localized techniques. However, recovering quantitative details about the identified phases which can be compared to traditionally measured metrics remains an enigmatic challenge. We demonstrate a technique to reconstruct phase-resolved magnetic hysteresis loops by selectively integrating the measured FORC distribution. From these minor loops, the traditional metrics – including the coercivity and saturation field, and the remanent and saturation magnetization – can be determined. In order to perform this analysis, special consideration must be paid to the accurate quantitative management of the so-called reversible features. This technique is demonstrated on three representative materials systems, high anisotropy FeCuPt thin-films, Fe nanodots, and SmCo/Fe exchange spring magnet films, and shows excellent agreement with the direct measured major loop, as well as the phase separated loops.

**Introduction**

Quantitatively distinguishing the coexistence of multiple magnetic phases from magnetometry studies has been a long-standing challenge in the magnetism community,[1] as the magnetic responses are often convoluted. For example, in heat-assisted magnetic recording (HAMR), ordered FePt alloys in the high magnetic anisotropy $L1_0$ phase is essential to allow for thermally stable ultra-small grain sizes and to increase the areal recording density.[2,3] The key challenge is to quantify the degree of ordering as the FePt is transformed from the as-deposited low anisotropy $A1$ phase to the $L1_0$ phase through thermal processing. Another example is in arrays of nanomagnets, the magnetization reversal mechanisms sensitively depend on the nanomagnet size, shape, and interactions, which makes it important to capture phase fractions of nanomagnets governed by different reversal mechanisms.[4,5] In more general cases of heterostructures consisting of magnetically hard/soft components that are exchange coupled, it is essential to distinguish the magnetic characteristics of each constituent.[6,7]

In this regard, the first-order reversal curve (FORC) method has gained popularity as a powerful approach to identify microscopic details of hysteretic systems through macroscopic measurements.[8-27] However, the correlation between the FORC distribution and traditional magnetometry is frequently non-trivial. Although the FORC technique can be used on virtually any hysteretic system with defined saturated states,[28-30] it is most commonly applied to magnetism.[31-33] While it is expected that the FORC distribution contains the traditional magnetometry metrics, such as coercivity and saturation fields, and the remanent and saturation magnetization, limited discussion exists on exactly how to capture these values. Furthermore, with the phase sensitivity of FORC, these values in principle can be determined for each phase. Developing an approach to reconstruct traditional magnetic measurements from within the often complex FORC distribution would both enhance the technique's capability, and promote a new level of comfort with the technique within the community.

In this work we demonstrate an approach to reconstruct phase-separated hysteresis loops from the FORC distribution. In order to perform an accurate reconstruction, the *full* FORC distribution must be considered, including the sometimes ignored and controversial reversible components.[34-41] This work starts by demonstrating approaches to managing reversible features, then demonstrates the reconstruction of the major hysteresis loops of three representative materials systems: (i) FeCuPt thin-films with mixed hard/soft phases for HAMR applications, (ii) Fe nanodots exhibiting both single domain and magnetic vortex configurations, and (iii) SmCo/Fe exchange spring magnet films with applications in permanent magnets. In each case the major loop is reconstructed and compared to the directly measured hysteresis loop; the loops are found to have excellent qualitative and even statistical agreement, validating our approach. Then, using the FORC technique to uniquely identify magnetic reversal behavior, phase separated hysteresis loops are reconstructed. This technique allows for the extraction of magnetometry metrics for individual phases within a multi-phase material, which is often difficult, even impossible, when considering major loops alone.

**Experiment**

In order to highlight the versatility of the FORC technique, three distinctly different types of systems are considered. The first example is a high magnetic anisotropy FePt based system, where a common challenge is the quantification of the $A1$ to $L1_0$ phase transformation. Here the $L1_0/A1$ mixed-phase $Fe_{39}Cu_{16}Pt_{45}$ thin films were deposited at room-temperature using an atomic multilayer sputtering technique, and subsequently treated by rapid thermal annealing (RTA), as reported previously.[42,43] For intermediate RTA temperatures at 300 °C - 400 °C, the $L1_0$ and $A1$ phases are exchange coupled at the microscopic level. Measuring these samples with the applied field in the out-of-plane (OOP) geometry, the $A1$ phase is reversibly forced OOP, while the (001) oriented $L1_0$ phase is hysteretic, thus representing an exchange coupled two-phase system with orthogonal easy axes. The second example is arrays of Fe nanodots with average diameters of 52 nm, 58 nm, and 67 nm, respectively, with edge-to-edge separation comparable to the dot diameter. They were grown by electron beam evaporation through alumina shadow masks and a subsequent lift-off process.[15,16,44] The smallest dots primarily undergo reversal by highly coherent rotation, while the largest reverse primarily by a vortex nucleation/propagation/annihilation mechanism. This transition occurs as a result of the geometrically dependent balance between the exchange and magnetostatic energies. The medium sized dots (58 nm) show both reversal mechanisms due to a finite size distribution, and represents a two-phase mixture system of weakly interacting elements. Lastly, the third example is an exchange spring thin-film MgO/Cr(20 nm)/Sm$_2$Co$_7$(20 nm)/Fe(10 nm)/Cr(5 nm) - henceforth referred to as SmCo/Fe - grown by magnetron sputtering.[12] As the applied magnetic field is first reduced from positive saturation to negative fields, the magnetically soft Fe film rotates reversibly with the field, while the magnetically hard SmCo remains in its initial orientation. At a sufficiently large negative field, the magnetically hard SmCo layer also switches; subsequently increasing the magnetic field, the 'winding' of the Fe occurs in positive fields due to its exchange coupling to the (now negatively saturated) SmCo.[6] This sample represents a two-phase system, with strong exchange coupling between the phases across the interface.

**FORC analysis**

Magnetometry measurements for all samples were performed on a vibrating sample magnetometer (VSM) at room temperature. FORC measurements were performed following procedures outlined previously:[8,10,35,45,46] From the positively saturated state the magnetic field is reduced to a scheduled reversal field, $H_R$, at which point the field sweep direction is reversed and the magnetization, $M$, is measured as the applied field, $H$, is increased to positive saturation, defining a single FORC branch. This process is repeated for $H_R$ between the positive and negative saturated states, thus capturing all reversible and irreversible behavior. For a regular field step of $\Delta H$, the $n+1^{th}$ FORC branch will begin at $H_R^{n+1} = H_R^n - \Delta H$. A mixed second order derivative is applied to the dataset to extract the normalized FORC distribution:

$$\rho(H, H_R) = -\frac{1}{2M_S}\frac{\partial}{\partial H_R}\left(\frac{\partial M(H,H_R)}{\partial H}\right) \quad \text{(Eqn. 1).}$$

Following the measurement sequence provided above, as $H$ is increased from an $H_R$ towards positive saturation, the FORC branch probes the up-switching field for all the magnetic elements that have been down-switched at $H_R$; up-switching events along this particular FORC are identified explicitly by the derivative *dM/dH*. Subsequently, by taking the $dH_R$ derivative, the new up-switching events on each FORC branch are isolated.[46] New up-switching events on the FORC branch starting at $H_R^n$ are not present on the branch starting at $H_R^{n-1}$, illustrating that the $H_R$-axis probes down-switching events. Thus the $H$ and $H_R$ axes separately identify the up- and down-switching events, but not necessarily of the same magnetic element; the FORC technique has no spatial resolution and cannot attribute up- and down-switching events to any particular element. This distinction motivates the term hysteron used to illustrate an elemental hysteretic event.[35] However, correlating hysterons to physical elements remains a challenge of the FORC technique.[23,46] The FORC diagrams can alternatively be represented in terms of a local coercivity and bias field: $H_B = \frac{H+H_R}{2}$, $H_C = \frac{H-H_R}{2}$.

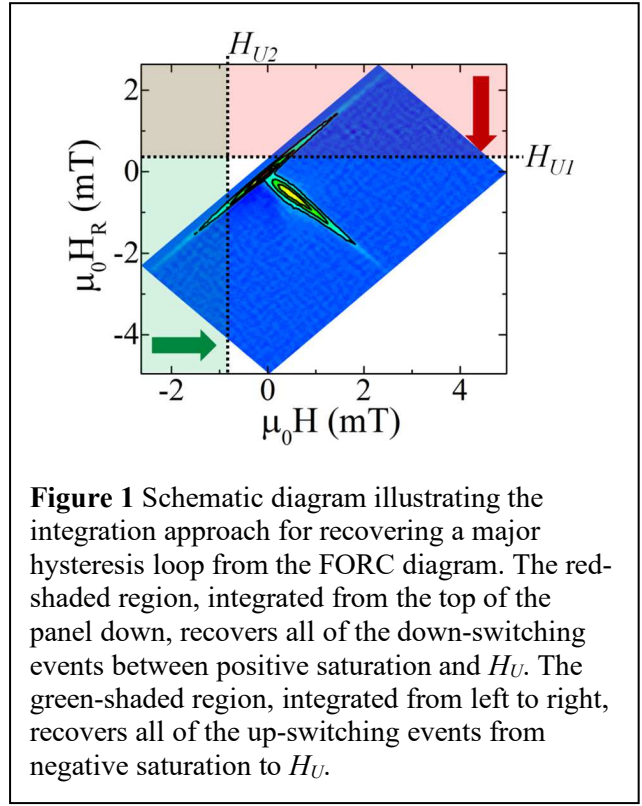

**Figure 1** Schematic diagram illustrating the integration approach for recovering a major hysteresis loop from the FORC diagram. The red-shaded region, integrated from the top of the panel down, recovers all of the down-switching events between positive saturation and $H_U$. The green-shaded region, integrated from left to right, recovers all of the up-switching events from negative saturation to $H_U$.

FORC projections are calculated by integrating the 2-dimensional FORC distribution over $H$ or $H_R$:
$$f(H_R) = \int_{-\infty}^{\infty} \rho(H, H_R) dH, \text{ and } f(H) = \int_{-\infty}^{\infty} \rho(H, H_R) dH_R \qquad \text{(Eqn. 2)}.$$

Thus, for example, $f(H_R)$ integrates all up-switching events that have down-switched at each $H_R$. Similarly, $f(H)$ integrates all down-switching events which would up-switch at each $H$. A subsequent integration of $f(H_R)$ or $f(H)$ over all $H_R$ or $H$, respectively, would capture all switching events and recover the saturation magnetization - assuming the reversible component is included in $\rho$.[34] In contrast, performing a partial integral, up to a dummy field variable called $H_U$, of the $H_R/H$ projection captures all the down/up switching events which have occurred *up-to* $H_U$. For example, $\Delta M(H_R = H_{u1} \to \infty) = \int_{H_{u1}}^{\infty} \int_{-\infty}^{\infty} \rho(H, H_R) dH \, dH_R$ integrates the red region in Fig. 1, and recovers the magnetization of all elements which down-switch between $H_{U1} < H_R < \infty$ (and up-switch between $-\infty < H < \infty$). By then subtracting $\Delta M$ from $M_S$, the magnetization along the descending-field branch of the major loop at $H_{U1}$ is recovered. Using a definite integral, e.g. spanning a defined parameter space, the magnetization along the descending branch of the major loop can be calculated:

$$M(H_{U1}) = M_S(1 - 2\int_{-\infty}^{\infty}\int_{H_{U1}}^{\infty} \rho(H, H_R) dH_R\, dH) \quad \text{(Eqn. 3).}$$

Similarly, integrating the $H$ projection over $-\infty < H < H_{U2}$, $-\infty < H_R < \infty$, identified as the green region in Fig. 1, calculates the magnetization of all of the up-switching events from negative saturation to $H_{U2}$. Using these values, the magnetization along the ascending branch of the major hysteresis loop can be determined:

$$M(H_{U2}) = -M_S(1 - 2\int_{-\infty}^{H_{U2}}\int_{-\infty}^{\infty} \rho(H, H_R) dH_R\, dH) \quad \text{(Eqn. 4).}$$

Carrying out calculations of magnetization at varying $H_{U1}$ ($H_{U2}$) between the saturated states, according to Eqns. 3 and 4, then recovers the descending and ascending-field branches of the major loop, respectively. The leading factor of $M_S$ in Eqns. 3 and 4 cancels with the $1/M_S$ prefactor in Eqn. 1, resulting in an absolute scale; for a normalized scale the prefactor in Eqns. 3 and 4 is one.

This integration technique becomes much more powerful when combined with FORC's capability to resolve multiple magnetic phases in a sample. That is, the FORC distribution has been demonstrated to uniquely identify behaviors within multiphase systems. Here, we first discern the separate magnetic phases in the as-measured FORC distribution and then employ Eqns. 3 and 4 to recover the phase-separated (major) hysteresis loops. While the bounds of Eqn. 3 and 4 are infinite and $HU_1$ and $HU_2$ – which can also span towards infinity – the reconstruction of hysteresis loops from individual features must integrate over a truncated range that includes only the FORC feature of interest. The bounds of this truncated space can be determined by the FORC distribution merging with the background signal; integrating further recovers no additional contributions to the magnetization since $\rho=0$ in these regions.

*Reversible Features*

Recovery of the embedded hysteresis loops often necessitates accurately accounting for the reversible features. However, reversible features manifest on the $H = H_R$ boundary of the dataset, where the FORC derivative is poorly defined. Specifically, the $n^{th}$ FORC branch has data spanning a field range $H_R^n \leq H \leq H_S$, where $H_S$ is the saturation field, and the neighboring $(n+1)^{th}$ FORC branch data spans $H_R^n - \Delta H \leq H \leq H_S$. Since there is no data at M($H = H_R^n - \Delta H, H_R = H_R^n$), which is required to calculate the FORC distribution at $(H = H_R^n, H_R = H_R^n)$, the derivative in Eqn. 1 is now ill-defined. Due to this incomplete data set at the $H = H_R$ boundary, reversible features – which by their very nature occur in this first field-step along a given FORC – are often not reported in the literature, and focus is instead placed on the irreversible components away from the $H = H_R$ boundary.

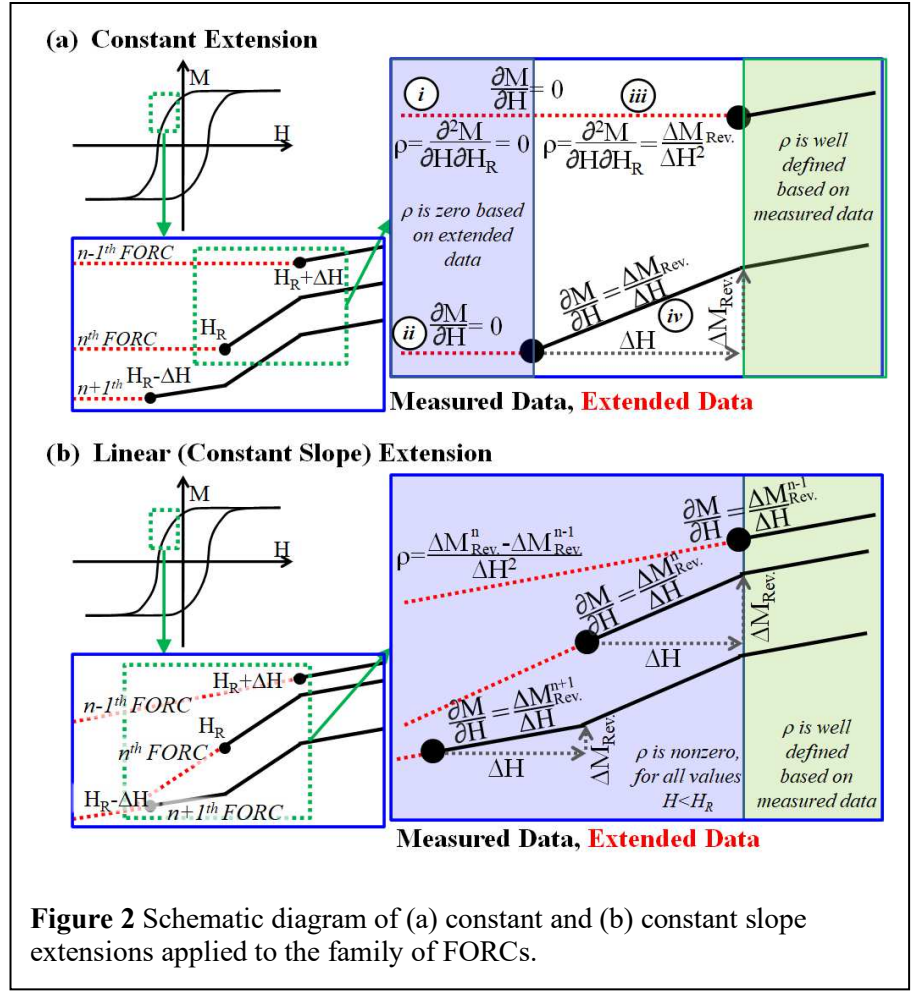

**Figure 2** Schematic diagram of (a) constant and (b) constant slope extensions applied to the family of FORCs.

Previous work by Pike[34] has demonstrated a constant extension scheme, graphically illustrated in Fig. 2(a), which recovers the reversible contribution to the magnetization. In this approach, the magnetization for $H < H_R$ is defined as: $M(H < H_R) \equiv M(H_R)$. Using this definition, $\frac{\partial M}{\partial H}\big|_{H<H_R} = 0$ in the extended region and Eqn. 1 simplifies to $\rho(H < H_R) = 0$. At the boundary, $\rho(H = H_R) = \frac{\Delta M_{Rev}}{\Delta H^2}$, where $\Delta M_{Rev}$ is associated with the reversible component of the magnetic reversal (See Fig. 2(a)). While there is solid analytical support for this approach, its quantitative application is rarely discussed in the literature.

An alternative approach is the constant slope - or linear - extension. For a linear extension $M(H < H_R)$ is defined such that $\partial M(H < H_R)/\partial H \equiv \partial M(H = H_R)/\partial H$, as illustrated in Fig. 2(b). Since $M(H < H_R)$ depends on both $H$ and $H_R$, this extension leads to FORC features throughout the extended data region ($H < H_R$). As a result, the integrated weight of the linearly extended FORC distribution will depend on the area over which the integration is performed. Below, we demonstrate that integrating the FORC distribution in ($H$, $H_R$) coordinates can recover the major loop saturation magnetization, while integrating in the transformed ($H_B$, $H_C$) coordinates can suppress the reversible phase.

A third approach most-often used to manage reversible behavior is to 'clip' the data-set. That is, to not calculate the FORC distribution in the region near the $H = H_R$ boundary of the dataset, until sufficient data at $H \geq H_R$ are available to determine $\rho$. In so doing the reversible contribution is rejected and does not contribute to the weight of the resultant distribution. Furthermore, clipping the dataset has the consequence of removing features which may be hysteretic but with a small coercivity, especially when relatively large magnetic field step sizes are used in the measurement, and thus is less-than ideal for certain samples with a significant fraction of magnetically soft phases. These three approaches are demonstrated on each of the samples in the results section below and on a fully-reversible YIG sphere calibration standard in the supplemental material, Figure S1a-c.

A physical picture based on Fig. 2 can be constructed and the integrated weight of the reversible FORC features calculated. The starting points on the $n$-1[th], $n$[th], and $n$+1[th] FORC have coordinates ($H$, $H_R$) of $(H_R^n + \Delta H, H_R^n + \Delta H), (H_R^n, H_R^n), and\ (H_R^n - \Delta H, H_R^n - \Delta H)$, respectively, marked by solid black dots. For the constant extension approach with $\frac{\partial M(H<H_R)}{\partial H} = 0$ on every FORC branch, as shown in Fig. 2(a),

$$\rho(H = H_R^n - \Delta H, H_R = H_R^n) \approx -\frac{1}{\Delta H}\left(\frac{\partial M}{\partial H}\bigg|_{H_R=H_R^{n-1}}^{H_R^n - \Delta H < H < H_R^n} - \frac{\partial M}{\partial H}\bigg|_{H_R=H_R^n}^{H_R^n - \Delta H < H < H_R^n}\right) = 0,$$

for the first step at $H < H_R$; these are labeled with *i* and *ii* in Figure 2a. In this region the extended data does not contribute to the FORC distribution. Also, the FORC distribution for $H > H_R$ is well defined and calculable by Eqn. 1. At the boundary of the dataset the FORC distribution can be calculated following Fig. 2(a). Specifically, the $(n-1)^{th}$ FORC branch starting at $H_R^n + \Delta H$ is shown with a constant extension, and the derivative calculated at the location *iii* is $\frac{\partial}{\partial H}M(H = H_R^n, H_R = H_R^{n-1}) = 0$. Then, applying Eqn. 1 between locations *iii* and *iv* identified in Figure 2a: $\rho \approx -\frac{1}{\Delta H}(0 - \frac{\partial M}{\partial H}\big|_{H_R=H_R^n}^{H_R^n < H < H_R^n + \Delta H}) = \frac{\Delta M_{Rev}^n}{\Delta H^2}$. This new feature quantitatively captures the reversible contributions on the $n^{th}$ FORC branch and manifests at $H = H_R^n$ (i.e. $H_C=0$). Repeating this for all the FORC branches and integrating the FORC distribution:

$$\int_{-\infty}^{\infty}\int_{-\infty}^{\infty} \rho\, dH dH_R = \int_{-\infty}^{\infty}\int_{H_R+\Delta H}^{\infty} \rho\, dH dH_R + \int_{-\infty}^{\infty}\int_{H_R}^{H_R+\Delta H} \rho\, dH dH_R + \int_{-\infty}^{\infty}\int_{-\infty}^{H_R} \rho\, dH dH_R$$

$$= M_S^{Irr} + \sum_n \frac{\Delta M_{Rev}^n}{\Delta H^2}\Delta H^2 + 0 = M_S^{Irr} + M_S^{Rev} = M_S^{Major\ Loop}$$

where the integral was re-written to reflect the discrete stepping performed in the FORC measurement. These integrals represent the region $H>H_R$, $H=H_R$ and $H<H_R$, respectively, e.g. the traditional FORC measurement, the reversible region, and the extended region which contributes zero. Thus, the analytical calculation shows integrating the constant extension accurately recovers the full saturation magnetization. This exercise also provides another useful insight: the constant extension to the dataset does not contribute to the FORC distribution except at the boundary.

Similar calculations can be performed on the linear extended data. Using Fig. 2(b) and Eqn. 1, the weight of the FORC distribution at any point in the extended region, including the boundary of the dataset, can be calculated: $\rho \approx \frac{1}{\Delta H}(\frac{\Delta M_{REV}^n}{\Delta H} - \frac{\Delta M_{REV}^{n-1}}{\Delta H})$. Then, integrating the FORC distribution in $(H, H_R)$ coordinates:

$$\int_{-\infty}^{\infty}\int_{Hq}^{\infty} \rho \, dH dH_R = \int_{-\infty}^{\infty}\int_{H_R}^{\infty} \rho \, dH dH_R + \int_{-\infty}^{\infty}\int_{Hq}^{H_R} \rho \, dH dH_R = M_S^{Irr} + \sum_n \frac{(\Delta M_{REV}^n - \Delta M_{REV}^{n-1})}{\Delta H^2} \Delta H^2 \left(\frac{H_R^n - H_q}{\Delta H}\right)$$

where $H_q$ is a dummy variable more negative than the negative saturation field ($H_q < -H_S$) defining the boundary of the integration space. Next, noting that neighboring FORC branches are separated in $H_R$ by $\Delta H$ in a regular FORC dataset, this can be expanded by applying the same discrete stepping discussed above:

$$\int_{-\infty}^{\infty}\int_{Hq}^{\infty} \rho \, dH dH_R = M_S^{Irr.} + \frac{1}{\Delta H}\left((H_R^0 - H_q)(\Delta M_{REV}^1 - \Delta M_{REV}^0) + (H_R^0 - \Delta H - H_q)(\Delta M_{REV}^2 - \Delta M_{REV}^1)\right.$$
$$+ (H_R^0 - 2\Delta H - H_q)(\Delta M_{REV}^3 - \Delta M_{REV}^2) + \cdots$$

$$= M_S^{Irr.} + \frac{1}{\Delta H}\sum_{n=1}^{Sat-1} \Delta H \Delta M_{REV}^n + (H_R^{Sat} - H_q)\Delta M_{REV}^{HSat} - (H_R^0 - H_q)\Delta M_{REV}^0 = M_S^{Irr.} + M_S^{Rev.}$$

where we have used the superscript to identify the FORC branch, with $0$ and $H_{Sat}$ identifying branches starting from the positively saturated state, and achieving negative saturation, e.g. at $H_R=H_{Sat}$. We also use the fact that, at saturation there is no reversible contribution, and thus $\Delta M_{REV}^{HSat} = \Delta M_{REV}^0 = 0$. Indeed, integrating the linear extended data in $H/H_R$ coordinates recovers the total saturation magnetization, with both the reversible and irreversible contributions. We also note that the variable $H_q$ drops out in the final solution; $H_q$ can thus be extended to negative saturation, integrating the entire parameter space without additional contributions to the integrated magnetization.

Lastly, integrating the linear extended FORC distribution in the ($H_C$, $H_B$) parameter space:

$$\int_{-\infty}^{\infty}\int_{-\infty}^{\infty} \rho \, dH_C dH_B = \int_{-\infty}^{\infty}\int_0^{\infty} \rho \, dH_C dH_B + \int_{-\infty}^{\infty}\int_{Hq}^{0} \rho \, dH_C dH_B = M_S^{Irr.} + \sum_n \frac{(\Delta M_{REV}^n - \Delta M_{REV}^{n-1})}{\Delta H^2}\left(\frac{\sqrt{2}H_q}{\Delta H}\right)\Delta H^2 =$$

$$= M_S^{Irr.} + \left(\frac{\sqrt{2}H_q}{\Delta H}\right)\left((\Delta M_{REV}^1 - \Delta M_{REV}^0) + (\Delta M_{REV}^2 - \Delta M_{REV}^1) + (\Delta M_{REV}^3 - \Delta M_{REV}^2) + \cdots\right)$$

$$= M_S^{Irr.} + \left(\frac{\sqrt{2}H_q}{\Delta H}\right)(\Delta M_{REV}^{HSat} - \Delta M_{REV}^0) = M_S^{Irr.}$$

As noted above, $\Delta M_{REV}^{HSat} = \Delta M_{REV}^0 = 0$ presuming the first and last FORC branches start and end in a saturated state, and there is no $H_q$ in the final result; integrating the linear extended FORC distribution in ($H_C$, $H_B$) coordinate system recovers only the irreversible component of the magnetization. This can be qualitatively visualized in the supplemental Figure S1d and can be conceptually understood as integrating the FORC feature with axes parallel to the reversible feature versus at 45°. This approach thus allows the irreversible features to be quantitatively separated without arbitrary clipping of the data, allowing powerful new insights to be made into mixed phase systems.

**Results**

*Reversible Feature*

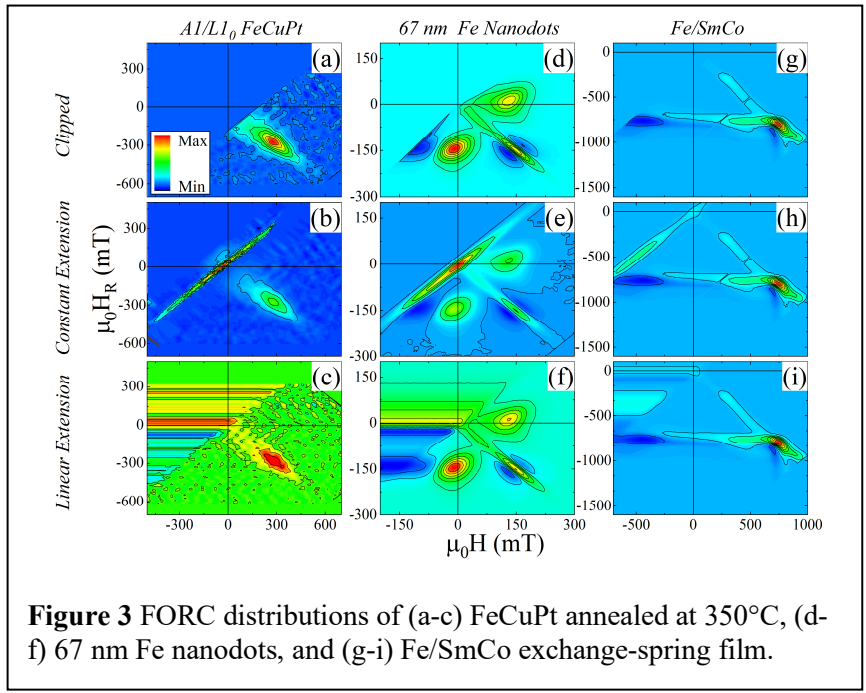

**Figure 3** FORC distributions of (a-c) FeCuPt annealed at 350°C, (d-f) 67 nm Fe nanodots, and (g-i) Fe/SmCo exchange-spring film.

FORC distributions for the FeCuPt annealed at 350 °C (Figs. 3(a-c)), 67 nm diameter Fe dots (Figs. 3(d-f)), and the Fe/SmCo exchange spring samples (Figs. 3(g-i)) were processed using clipping (top row), constant (middle row) and linear (bottom row) extensions. As discussed above, the clipped data entirely removes the reversible feature, while the constant extension method reveals a reversible feature only at $H = H_R$. Integrating the clipped data recovers an approximate ratio for the irreversible magnetization vs. major loop saturation magnetization $M_S^{Irr}/M_S^{Major\ Loop}$ of (3a) 0.28, (3d) 0.53, and (3g) 0.67. By comparison, integrating the constant extension distributions (middle row of Fig. 3) accurately recovers the major loop saturation, with $M_S^{FORC}/M_S^{Major\ Loop}$ of (3b) 1.03, (3e) 0.98, and (3h) 0.94.

Comparing the above values from the clipped and constant extension data confirms the above derivations; similar analysis confirms the constant-slope extension. Integrating the FORC distribution with a constant-slope extension in $(H_C, H_B)$ recovers a magnetization of $M_S^{FORC}/M_S^{Major\ Loop}$ =(3c) 0.35, (3f) 0.53, and (3i) 0.55, in reasonably good agreement with $M_S^{Irr.}$ as measured by clipping. Integrating the FORC distribution processed with a constant-slope extension in $(H, H_R)$ coordinates accurately recovers the major loop saturation magnetization: $M_S^{FORC}/M_S^{Major\ Loop}$ = (3a) 0.9, (3b) 1.0, (3c) 0.9. Thus, we confirm good agreement between the clipped dataset and the linear extension in $(H_C, H_B)$ coordinates, as well as between the constant extension and the linear extension in $(H, H_R)$ coordinates. It is less clear what additional information is recovered by integrating the linear extended data (bottom row of Fig. 3).

A more definitive demonstration is shown in the FORC measurements of a YIG sphere (Supplemental Materials). The YIG sphere standard has a closed hysteresis loop, showing fully reversible behavior. Integrating the FORC distribution processed with the different extensions recovers the major loop $M_S$ or zero.

In order to reconstruct the major loop the entire magnetization must be recovered and the constant extension or linear extension (integrated in $(H, H_R)$) must be utilized. However, when we consider applying the integrated area represented by the red box in Fig. 1, to the constant-slope extension, shown in Fig. 3(c, f, i), it is clear that the integrated magnetization will depend on the integration limits in $-H$ (analogous to the variable $H_q$ discussed above). For this reason, to reconstruct the major loop the constant extension is the reasonable choice.

***Constructing Hysteresis Loops***

*Mixed Phase FeCuPt*

FeCuPt samples are annealed by RTA at 300 °C - 400 °C, resulting in a mixed soft/hard $A1/L1_0$ phase. The FORC distributions, Fig. 4(a-e), show the two-phase system consisting of a mostly reversible $A1$ phase (located at $H_C=0$) and an irreversible $L1_0$ phase (located at finite $H_C$). Increasing the annealing temperature (left to right in Fig. 4) leads to a decrease in the weight of the reversible phase, consistent with the transformation from the $A1$ to $L1_0$ phase. In the previous section the constant extension was shown to recover the total saturation magnetization, as measured from the major loop, and will be used here to reconstruct the major loop.

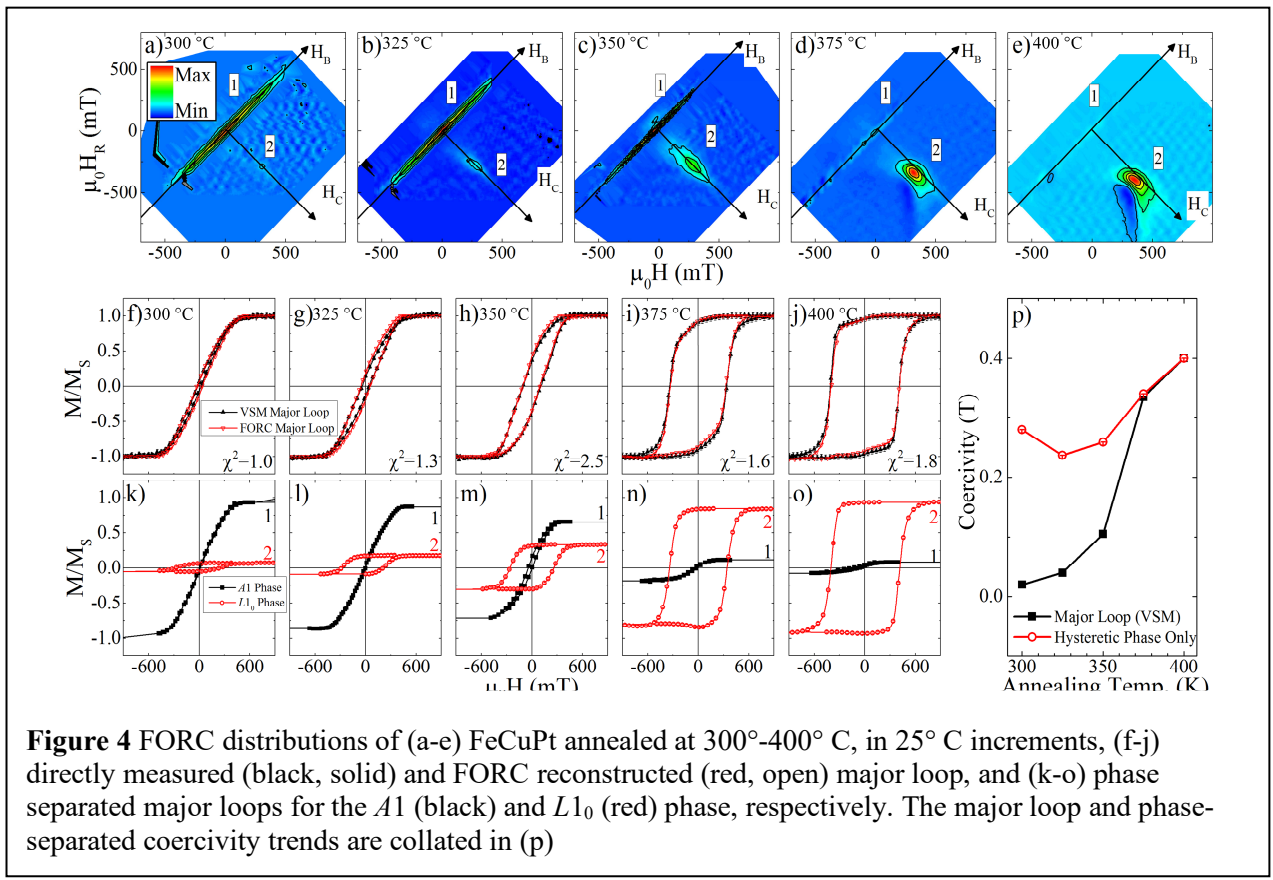

**Figure 4** FORC distributions of (a-e) FeCuPt annealed at 300°-400° C, in 25° C increments, (f-j) directly measured (black, solid) and FORC reconstructed (red, open) major loop, and (k-o) phase separated major loops for the $A1$ (black) and $L1_0$ (red) phase, respectively. The major loop and phase-separated coercivity trends are collated in (p)

Performing the progressive integral of the FORC distribution along $H$ and $H_R$ was suggested above to recover the ascending and descending branches of the major loop (Eqn. 3 and 4), respectively. The reconstructed major loops are shown superimposed with the major loops as measured directly with the VSM in Fig. 4(f-j). The reconstructed and directly measured major loops show excellent agreement. Quantitative agreement between the loops is further shown by evaluating the difference between the loops and calculating the reduced $\chi^2$ comparing to a null-set model (i.e. the model tests if the two loops are statistically identical). The error for each measurement is defined by the machine sensitivity of 3 μemu. For every case the reduced $\chi^2$ is listed in the figure panel and is approximately unity, confirming the excellent agreement within a characterized error.

Next, each of the two primary features in the FORC distribution are isolated and integrated to recover the phase-separated hysteresis loops, Fig. 4(k-o). Integrating the feature at ($H = H_R$), labeled 1 in the FORC diagram, recovers a major loop confirming its role as the reversible phase which diminishes with increased RTA temperature. The hysteretic phase, labeled 2 in the FORC diagrams, shows an open loop with increasing magnetization and coercivity with increasing RTA temperature. Comparing these loops and the directly measured hysteresis loops emphasizes the need to include the reversible feature in the above full-loop reconstruction. The coercivity for the hysteretic phase is extracted and plotted in comparison to the major loop coercivity in Fig. 4(p). At the higher RTA temperatures, the hysteretic phase constitutes the majority of the sample and the coercivities agree well. However, as the $L1_0$ phase fraction is reduced at lower RTA temperatures, coercivities from the major loop and the extracted $L1_0$ phase diverge. The extracted $L1_0$ loop identifies that the coercivity of the $L1_0$ phase levels off at approximately 250 mT at low annealing temperatures (red curve in Fig. 4p), even when the FeCuPt is primarily in the $A1$ phase. This revelation provides sharp contrast to that from the measured major loop (black curve in Fig. 4p), which shows $H_C$ roughly scales with annealing temperature. The reconstructed $L1_0$ loop thus provides crucial insights to the ordering mechanism of the $A1$ to $L1_0$ phase transformation: this is consistent with the nucleation/growth model of the $L1_0$ phase, which suggests that each grain should transform from $A1$ to the high anisotropy $L1_0$ phase and is therefore expected to have a reasonable coercivity, once $L1_0$ ordering occurs.

*Single Domain and Vortex State Fe Nanodots*

FORC measurements of Fe nanodots, with average diameters of 52 nm - 67 nm, were one of the earliest applications for fingerprinting different reversal mechanisms and quantitative phase determination.[15,16] The FORC diagrams, Fig. 5(a-c) show the evolution from a single domain state, 52 nm diameter dots in panel (5a), to vortex-state reversal, 67 nm diameter dots in panel (5c). The intermediate sized dots, 58 nm diameter dots in panel (5b), show the presence of both reversal mechanisms. A reversible feature is expected during vortex reversal, as the vortex core moves reversibly inside the nanodot in response to an external magnetic field. Interestingly, all three samples show a significant feature at $H_C=0$ using the constant-extension approach; the extent of the reversible feature along $H_B$ beyond the vortex annihilation feature and its presence in the 52 nm sample indicates another reversible component. While this reversible component was not addressed in the original study, some clues can be elucidated from the major loop, which clearly shows a reduced remanence, $M_R/M_S \approx 0.6$. Note that these samples were prepared by evaporation through a self-assembled template mask and exhibit a finite distribution of shapes and sizes.[15,16,44] This reversible component may represent a relaxation of the magnetization along the dot periphery preceding switching or perhaps even a distribution of easy axes in the plane.

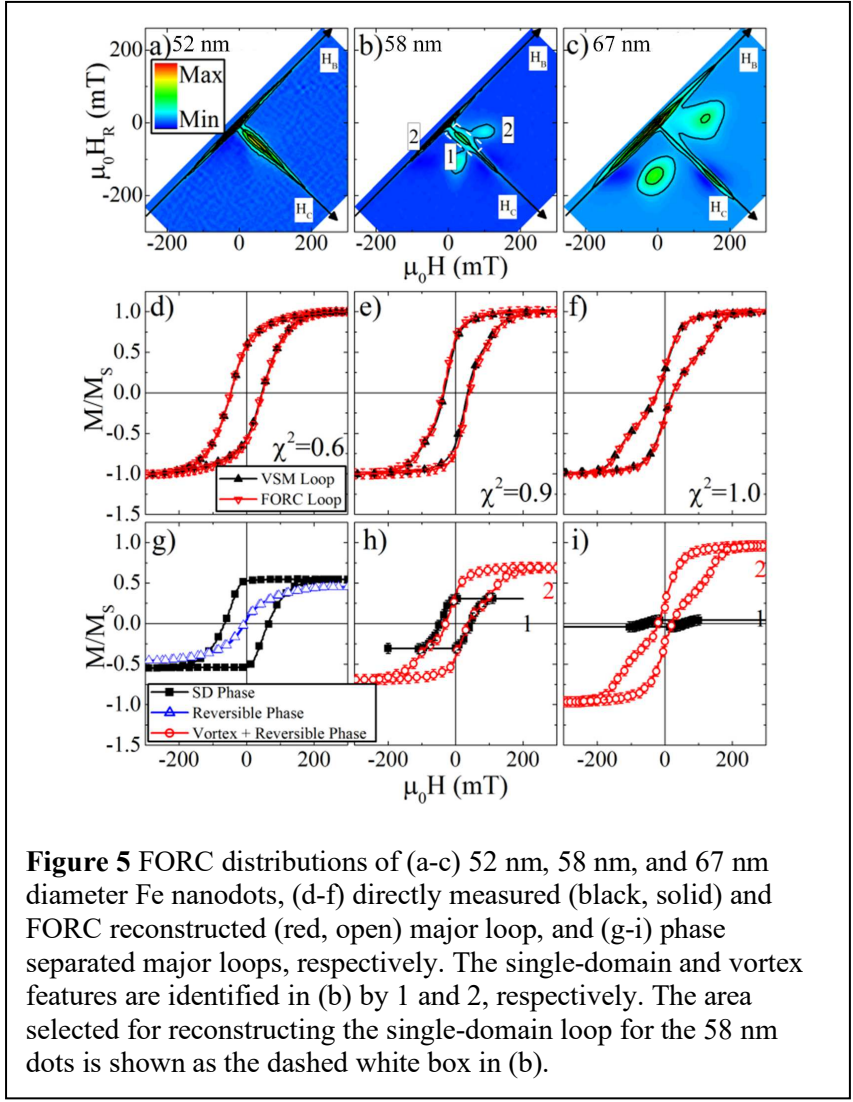

**Figure 5** FORC distributions of (a-c) 52 nm, 58 nm, and 67 nm diameter Fe nanodots, (d-f) directly measured (black, solid) and FORC reconstructed (red, open) major loop, and (g-i) phase separated major loops, respectively. The single-domain and vortex features are identified in (b) by 1 and 2, respectively. The area selected for reconstructing the single-domain loop for the 58 nm dots is shown as the dashed white box in (b).

Similar to the FeCuPt samples, the major loops are reconstructed by progressively integrating the FORC distribution, and compared to the VSM major loops in Fig. 5(d-f). Again, the reconstructed major loops show excellent qualitative and quantitative agreement with the directly measured major loops. Furthermore, using the FORC diagram to separate the reversal mechanisms, the hysteresis loop for each phase is reconstructed in Fig. 5(g-i). For the 52 nm dots the single domain and reversible phases are plotted and, similar to the FeCuPt sample, the coercivity for the hysteretic phase (66 mT) is significantly larger than the VSM major loop (46 mT). From the saturation magnetization of each phase the reversible and hysteretic phases contribute similarly (45% v. 55%) to the ensemble magnetism of the sample.

For the 58 nm and 67 nm dots, which possess three phases (single domain, vortex state, and reversible), only the single domain phase can be separated, since the vortex-state and reversible contributions have overlapping features at $H_C$=0. The single domain phase is shown to decrease in its contributions with increased dot size, and possess a coercivity always much larger than the VSM major loop. The vortex+reversible phase, on the other hand, develops into a wasp-waisted hysteresis loop and increases in weight with dot diameter, following its increased vortex phase fraction.

*Fe/SmCo Exchange Spring*

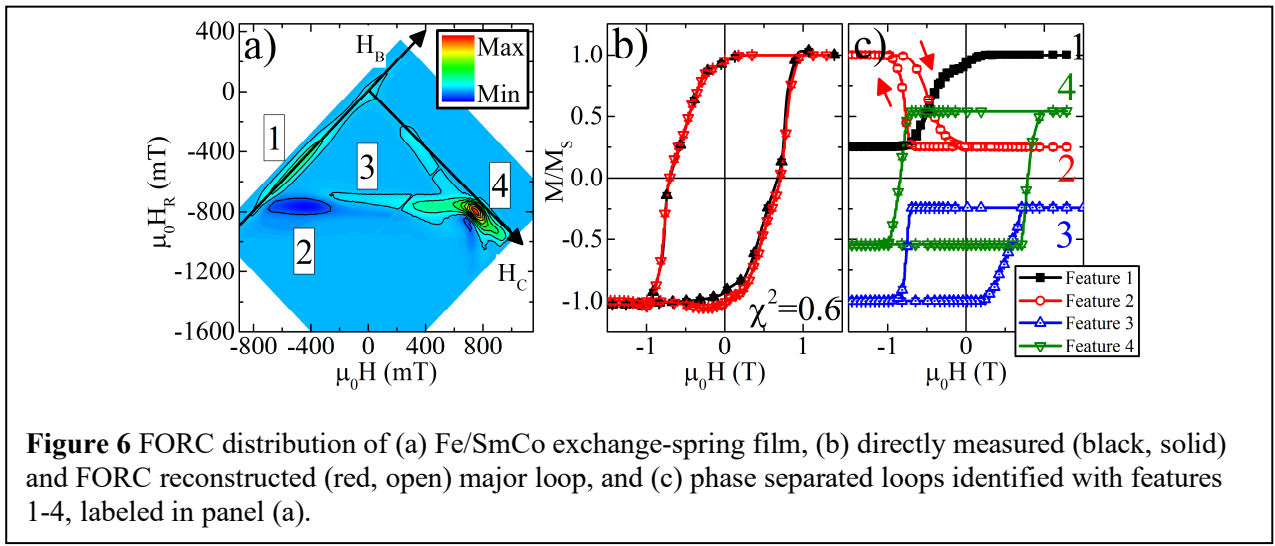

**Figure 6** FORC distribution of (a) Fe/SmCo exchange-spring film, (b) directly measured (black, solid) and FORC reconstructed (red, open) major loop, and (c) phase separated loops identified with features 1-4, labeled in panel (a).

The last example is a Fe/SmCo exchange spring thin-film shown in Fig. 6. As the magnetic field is reduced from positive saturation to small negative fields, the SmCo remains positively saturated while the Fe layer follows the field reversibly, which then relaxes to a positive orientation in the absence of a field due to the exchange coupling to the SmCo. This reversible 'winding' of the exchange spring occurs when the applied field is anti-parallel to the SmCo layer and manifests as Feature 1 in Fig. 6(a). Then, at sufficiently large negative fields the SmCo reverses. Along the subsequent FORC branch, the reversible winding of the Fe layer occurs now in positive fields – since the anchoring SmCo is now negatively oriented. Thus, along this FORC branch there are three magnetic features: First, across the applied field range for the initial winding (Feature 1) the magnetization versus field is now flat (SmCo and Fe are parallel and aligned with the magnetic field in the negative direction), thus dM/dH over this range decreases, from a positive value to zero, with decreasing $H_R$. Following from Eqn. 1, this generates a negative FORC feature (Feature 2); the origin of negative features was discussed in previous works.[46] Continuing along this FORC branch, in positive magnetic fields the Fe reversibly winds, with the SmCo still in the negative direction. This winding was absent on more-positive $H_R$ branches, and so constitutes an increase in dM/dH with decreasing $H_R$, generating a positive feature in the FORC distribution (Feature 3). Finally, at large positive fields, the SmCo reverses, generating Feature 4. The alignment in $H$ of Features 1 and 2 and alignment in $H_R$ of Features 2 and 3 attests to their common origin in the Fe layer.

The reconstructed major loop, shown in Fig. 6(b), once again shows excellent qualitative and quantitative agreement with the directly measured major loop. Phase separated major loops were calculated for each feature labeled 1-4 in Fig. 6(a), as shown in Fig. 6(c). While the other FORC distributions had well-separated phases, for the SmCo/Fe phases, features 3 and 4 are all overlapping; the integrated weight of feature 1 is used to set the limit on the integration for feature 3, thus separating the phases. Consistent with the explanation above, the loop from Feature 1 is closed, identifying the reversible winding of the Fe coming from positive saturation. Interestingly, the loop for Feature 2, which comes from a negative FORC feature, results in an inverted hysteresis loop. The descending branch of loop 2, e.g. from positive to negative magnetic fields, has a sharp switching event at the SmCo switching field, then the ascending branch runs parallel to the Feature 1 loop. The inverted correlation between the Feature 1 and 2 loops is a visual demonstration of the disappearance of the Fe winding in negative fields after the SmCo reversal. The loop from Feature 2 also demonstrates that the FORC technique maps model hysterons and not physical magnetic elements; the Feature 2 loop identifies the weight of the Fe layer, with an up-switching of the SmCo reversal field, and a down-switching reflective of the Fe winding (coming from positive saturation). Similarly, the loop from Feature 3 shows a down-switching at the SmCo reversal field, and up-switching as the Fe winds in positive fields, coming from negative saturation. Thus, between these three loops, all of the behavior of the Fe is captured. The SmCo loop is also extracted, and gives values for the coercivity which are consistent with the major loop value.

**Conclusion**

The FORC technique has become widely used in the magnetics community as a tool for identifying the microscopic details of often complex systems through macroscopic measurements. A particular strength of the FORC technique is its ability to uniquely identify and separate magnetic reversal behavior in multi-phase systems. Within the magnetic "fingerprints" of each phase are details of the magnetization reversal which can be recovered through appropriate numerical treatment. Prerequisite to this treatment was proper consideration of the reversible phase. Constant and linear data extensions were analytically and empirically demonstrated. Using a constant data extension, the major hysteresis loop was reconstructed from the FORC distribution. Then, using the FORC technique to uniquely identify magnetization reversal behaviors, phase separated hysteresis loops were reconstructed. These phase-resolved loops contained the traditional metrics of magnetic reversal behavior, including the coercivity and saturation field, and the remanent and saturation magnetization (which can be correlated to phase fraction). This work addresses traditional issues extant with the FORC technique - namely reversible features - and also promotes a new approach, which builds a bridge between the FORC technique and traditional magnetometry.

**Acknowledgments**

This work has been supported by the NSF (ECCS-1611424 and ECCS-1933527) and DOE Career (DE-SC0021344).